%% file: 0_main.tex
\documentclass[sigconf,dvipsnames]{acmart}
\renewcommand\footnotetextcopyrightpermission[1]{} % removes footnote with conference information in first column

\AtBeginDocument{%
  \providecommand\BibTeX{{%
    \normalfont B\kern-0.5em{\scshape i\kern-0.25em b}\kern-0.8em\TeX}}}

\setcopyright{acmcopyright}
\copyrightyear{2022}
\acmYear{2022}
\acmDOI{10.1145/1122445.1122456}

\acmConference[ICPC 2022]{The 30th International Conference on Program Comprehension}{May 21–22, 2022}{Pittsburgh, PA, USA}
\acmPrice{15.00}
\acmISBN{978-1-4503-XXXX-X/18/06}

\include{macro}
\begin{document}

\title{Recommending Bug Assignment Approaches for \\Individual Bug Reports: An Empirical Investigation}

% authors will be added in this order: Yang, Junayed, Oscar, Kevin, Andi, Denys
\author{Yang Song}
\email{ysong10@wm.edu}
\affiliation{%
	\institution{College of William \& Mary}
	\country{Williamsburg, Virginia, USA}
}
	
\author{Oscar Chaparro}
\email{oscarch@wm.edu}
\affiliation{%
	\institution{College of William \& Mary}
	\country{Williamsburg, Virginia, USA}
}

\renewcommand{\shortauthors}{Song et al.}

\begin{abstract}

Multiple approaches have been proposed to automatically recommend potential developers who can address bug reports. These approaches are typically designed to work for any bug report submitted to any software project. However, we conjecture that these approaches may not work equally well for all the reports in a project. We conducted an empirical study to validate this conjecture, using three bug assignment approaches applied on 2,249 bug reports from two open source systems. We found empirical evidence that validates our conjecture, which led us to explore the idea of identifying and applying the best-performing approach for each bug report to obtain more accurate developer recommendations. We conducted an additional study to assess the feasibility of this idea using  machine learning. 
%We experiment with multiple ML models that learn from 23 features that encode possible relationships between bug reports and the approaches. 
While we found a wide margin of accuracy improvement for this approach,
%by perfectly recommending the best approach for individual bug reports, 
it is far from achieving the maximum possible improvement and performs comparably to baseline approaches. We discuss potential reasons for these results and conjecture that the assignment approaches may not capture important information about the bug assignment process that developers perform in practice. The results warrant future research in understanding how developers assign bug reports and improving automated bug report assignment.

\end{abstract}

%uncomment this for the camera-ready



%\keywords{Bug reporting, Android apps, Task-based Chatbots}

\maketitle

%-------------------------------------

\input{1_intro}

\input{2_study_1}

\input{3_study_2}

\input{5_threats}
\input{6_related_work}

\input{7_conclusions}

%\input{data}

%-------------------------------------

\balance
%\todo{Check for repeated references and make the format of references consistent (author names, venues and their acronyms, broken links, etc)}

%\os{Check for repeated references, inconsistent formatting, and other issues in the references.}
\input{reference.bbl}

\bibliographystyle{ACM-Reference-Format}
\bibliography{references}

\end{document}

%% file: macro.tex
\usepackage[utf8]{inputenc}
\usepackage{url}
\usepackage{colortbl}
\usepackage{caption}
\usepackage{balance}
\usepackage{xspace}
\usepackage[caption=false]{subfig}
\usepackage{graphicx}
\usepackage{verbatim}
\usepackage{bold-extra}
\usepackage{amsmath}
\usepackage{multirow}
\usepackage{listings}
\usepackage{boxedminipage}
\usepackage{soul}
\usepackage{enumitem}
\usepackage[normalem]{ulem}
\setlist{nolistsep,leftmargin=.5cm}
\usepackage{booktabs}
\usepackage{tabularx}
\usepackage{svg}
\usepackage{calc}
\usepackage{float}
\usepackage{multirow}
\usepackage[normalem]{ulem}
\useunder{\uline}{\ul}{}
\usepackage{amsmath}
\usepackage[most]{tcolorbox}
\usepackage{caption}
\usepackage{float}
\usepackage{fancyhdr}
\newboolean{showcomments}
\setboolean{showcomments}{true}

\ifthenelse{\boolean{showcomments}}
  {\newcommand{\nb}[2]{
    \fbox{\bfseries\sffamily\scriptsize#1}
    {\sf\small$\blacktriangleright$\textit{#2}$\blacktriangleleft$}
   }
   
  }
  {\newcommand{\nb}[2]{}
   
  }

\newcommand{\ie}{\textit{i.e.},\xspace}
\newcommand{\eg}{\textit{e.g.},\xspace}
\newcommand{\etc}{\textit{etc.}\xspace}

\newcommand{\aka}{\textit{a.k.a.}\xspace}

\usepackage{tikz}

\definecolor{bug_red}{rgb}{.84,.23,.29}

\definecolor{info-needed-color}{rgb}{1,.8,.12}

%% file: 1_intro.tex
%\os{Let's use "approach" consistently throughout the paper}
\section{Introduction}
%\os{My impression of the introduction: the information is there but it is not written nicely. We should write it similarly to the abstract but giving more details (for example, what do we mean by bug report nuances?).}

%\os{This first paragraph should be short. The message should be something like: many software projects receive hundreds of bug reports on a daily basis that need to be triaged. In this process, one of the first steps is to assign them to the right developers, those with the  expertise to solve the bugs. This step can be time consuming and researchers have proposed techniques that automatically recommend potential developers based on the received bug reports and other information. These are intended to facilitate the task and reduce the effort of developers.}

%\os{The intro needs multiple passes, it needs to sound stronger}
Many software projects receive hundreds of bug reports on a daily basis that need to be triaged and solved as soon as possible \cite{zou2018practitioners,Anvik2005,Sun2011}. An important step in the bug triage process is to assign the bug reports to the developer who has the proper expertise to solve the reported bugs. Given the high volume of incoming reports, this process is often time-consuming \cite{zou2018practitioners,Xia2017,Anvik2005,Sun2011}. Hence, researchers have proposed approaches to automatically recommend potential developers for a bug report \cite{aktas2020automated,Anvik2006,hossen2014amalgamating,Kagdi2012,Kevic2013,Matter2009,Tamrawi2011a,Tian2016a,xia2015dual,Xia2017}. These approaches implement a variety of techniques, \eg machine learning (ML) \cite{Bhattacharya2012} or information retrieval (IR) \cite{zanjani2015using},
%from machine learning (ML) models \re and information retrieval (IR) techniques \re to topic models \re and tossing-graph-based techniques \re
 and leverage bug report information and data such as past reports/code changes, and the developers who processed these artifacts. %involved in processing these artifacts.

Existing bug assignment approaches use different artifact information and ways to encode such information for automatically recommending potential developers. More importantly, these approaches are typically designed to work for any bug report submitted to any software project. However, we conjecture that these approaches may not work equally well for all the reports in a software project, given their internal recommendation mechanisms and particular nuances that each bug report has (\eg the bug type and specific information that users report for such a type).

In this paper, we first report an empirical study (\aka Study~1) that aims to validate such conjecture. We executed three bug report assignment techniques,
%(namely, Learning to Rank or L2R \re, Lucene \re, and Frequency \re) 
that use distinct techniques (\eg ML or IR) and information from various software artifacts, 
on 2,249 bug reports from two open-source systems. We compared how often the three approaches would perform best in recommending the expected developers for the reports (based on ground-truth data).
%\os{We need to give numbers here} 
We found that no single approach performed best for the majority of the bug reports in each system. Instead, each of the three approaches performs better for some reports and worse for others in both software systems, which means that these approaches do not work equally well for all the bug reports in a project.
%\os{I think we shouldn't not explain in detail the approaches here. A brief sentence for each should suffice at this point.}
%L2R is one state-of-art approach that applies learning-to-rank algorithm to rank the most suitable developers to fix a bug report. Lucene is based on bug report similarity, that is, to recommend developers who solve the similar bug reports; Freq recommends the most frequent developers who solve bugs in the past. 
%Our results show that no models perform the best for all projects, in particular, these approaches do not work equally well for all the reports in a project.
% L2R is one state-of-art approach  which is a machine learning technique that map the assign recommendation task to a document retrieval task by treating the bug report as query, and developers' profiles as documents to be returned. It uses  potential location of the bug (location-based features) and previous fixed bugs by each developer (activity-based features) to represent a developer's profile and train learning to rank model to learn weights for features and finally recommend developers according to the ranking scores;
%\os{I like the next sentences. May we should say  that because different methods consider different types of info and work differently,  it makes sense to combine them by recommending them to individual bug reports, so that we exploit their advantages (sometime like that)} 
%\yang{this is already mentioned before}
The results motivated us to explore the idea of automatically identifying the best approach, from the three ones used in Study~1, for each individual bug report and running such an approach to produce more accurate developer recommendations. In this way, we would combine the advantages of the three approaches to produce better recommendations. To the best of our knowledge, this is the first work that explores this idea for automated bug assignment.

%These results inspired us to design a composite approach that combines the advantages of three bug assignment approaches and determine the most suitable approach for each bug report to obtain more accurate developer recommendations. 

We report an additional study (\aka Study 2) to assess the feasibility of this idea by using machine learning (ML). We experimented with multiple ML models to recommend the best-performing bug assignment approach for each bug report. These models are trained and evaluated using the same bug report dataset from Study~1 and 23 features that encode possible relationships between the reports and the approaches. Once the models detect the best performing approach for a report, this is applied  on the report to obtain a list of potential developers who can address the reported bug. We call this approach \textit{Lupin}. The results show that, while there is a wide margin of accuracy improvement by perfectly recommending the best approach for individual  reports, \textit{Lupin} is far from achieving the maximum possible improvement and performs comparable to using the individual assignment approaches on all the reports. We discuss potential reasons for these results and conjecture that existing approaches may not capture important information about the bug assignment process that developers perform in practice. The results warrant future research in understanding how developers assign bug reports and improving automated bug report assignment.

%% file: 2_study_1.tex
\section{Study 1}
%\section{Bug Assignment on Single Bug Reports}
%\os{We need a better title for this section. I would avoid using bug triaging/triage, except the first paragraph of the intro.}

%\os{I cannot think of a good, short title for this section. Maybe in the intro we label the studies as "Study 1" for the one about validating the conjecture, and "Study 2" for assessing the idea of recommending the approaches to individual bug reports. Then, we use these labels as section titles}
We conducted an empirical study to validate the conjecture that existing bug report assignment techniques do not work equally well for all the bug reports in a software project. Specifically, the study aimed to answer the following research question:

\begin{center}
    \textit{\textbf{RQ$_1$}: How do bug report assignment approaches perform and compare when applied on individual bug reports?}
\end{center}

%To answer RQ$_1$, we first collected bug report data for two open source projects. Then, we applied three existing bug assignment approaches on the collected reports. We measured the performance of the approaches and the distribution of bug reports for which each approach gave the best developers recommendation.

\begin{comment}

To answer RQ$_1$, we collected bug reports and associated data (\eg developers who solved the bugs) for two open source projects (Sect. \ref{sec:dataset}). Then, we selected and implemented three existing bug assignment approaches  and applied them on the collected reports (Sect. \ref{sect:approaches} and \ref{sec:metrics}). We measured the performance of the approaches using well-known metrics and the distribution of bug reports for which each approach gave the best developers recommendation (Sect. \ref{sec:metrics}). We discuss the results to answer the RQ in Sect. \ref{sec:results}.
\end{comment}

% \subsection{Approach Overview} 
%\os{We don't need this subsection. Instead we need here to introduce the research question (RQ1). Then, here, after we introduce the RQ, in couple sentences we explain the overal methodology to answer the RQ, which is expanded on each of the subsections below.}

\subsection{Dataset}
\label{sec:dataset}

%\os{We need to say somewhere here why did we pick Angular and Wordpress Android? What it makes sense to use these systems for this study? We need to say that they are used in prior studies (cite the papers), the are mature and large systems, of different types of software, for different domains, etc.}
%\yang{check the paper that use these datasets}
We collected bug report data for two open source projects, namely Angular.js (Angular) and WordPress Android (WordPress), which have been used in prior work \cite{Sajedi2020,Chaparro:FSE17}. Angular is a web framework and WordPress is an Android app for website creation. These projects were selected because they are active, large (90kLOC+), support different domains, and involve a large number of developers (171 and 40 for Angular and WordPress, respectively). 
%These systems have also been used in prior work on bug report assignment \re and issue quality verification \re.

%As these projects are hosted on GitHub, 
We used the GitHub API to download all of the project issues submitted until Oct. 2020. For each issue, we collected the ID, the title/summary and description, the status, the date-time of submission, 
%the date-time when the issue was closed, 
the labels, and the person who was assigned to solve the issue. 
%Since not all of the issues on GitHub describe bugs for the systems, 
We used the issue labels to identify the bug reports: those issues tagged with the label ``bug'' by the project maintainers. We only used those reports that were closed and had one or more commits that fixed the bug (read below). We collected 905 and 1,344 reports for Angular and WordPress, respectively (2,249 reports total).

As in prior work \cite{Sajedi2020,hossen2014amalgamating}, 
%based on bug report data, 
we built the set of developers who fixed the bug for each bug report (\ie the ground truth). 
%These sets represent the ground-truth for automated bug report assignment. 
We consider the developer assigned to solve the bug on the issue tracker as well as the developers who pushed the code changes to solve the bugs as the ground-truth developer set for each report. To build the developer sets, we identified the associated commits with each bug report by using regular expressions on the commit and bug report comments. From the identified commits, we extracted the developers who authored or committed the code changes. 
%We searched for  phrases in the commit messages that matched patterns such as "fix/fixed/fixed [ISSUE\_ID]" \cite{github-close-issues}. We extracted the [ISSUE\_ID] from the commit messages to identify the corresponding bug report and link it to the commit. We also used  a similar approach to search for commit ID/hash references in the report comments.  From the identified commits, we extracted the developers who authored or committed the code changes. 
%Since the same bug assignee and/or commit developer may use different GitHub/Git accounts, we manually merged the accounts that referred to the same person. 
The avg. number of developers that fixed the bug in the collected bug reports is 1.4 (Angular) and 1.1 (WordPress). We applied tokenization, stop word removal, and stemming on all the textual data (\eg bug reports).

\subsection{Bug Assignment Approaches}
\label{sect:approaches}
% \textit{ How do different bug triaging methods perform on different systems?}

%\os{add citations here and explain more clearly how L2R is able to recommend developers}

We used different types of bug assignment approaches that leverage information from various artifacts to recommend potential developers. We used an ML-based approach (L2R \cite{Tian2016a}), an IR-based approach (Lucene), and another approach that leverages how frequent the developers solve bugs (Freq). These approaches recommend a ranked list of developers for a particular bug report (\aka the query, formed by concatenating its title/summary and description). Developers ranked higher in the list are considered to have more expertise in addressing the report.

%\os{TODO: reduce this paragraph}
\textbf{L2R} is a state-of-the-art approach \cite{Tian2016a} (based on the Ranking SVM model) that learns to rank tuples composed of a bug report and developer, based on 16 features that represent the similarity/relevance between the report and the developer. %The features are categorized within two groups, namely activity-based and location-based. The activity-based features encode information related to the bug fixing activities that developers perform in a project.
%and are extracted from various artifacts such as past bug reports and code files related to a given report and developer. 
For example, a set of features leverage the textual similarity between the bug report and the code files that a developer modified (via VSM \cite{Salton1975} and BM25 \cite{Robertson2009}). 
%The location-based features represent information related to code locations that potentially contain the bug described in the report. 
Another set of features leverage the textual similarity between potentially-buggy code files (as given by a bug localization approach, based on the report) and the code files modified by the developer. 
%L2R uses RankSVM \re to learn the weights (importance) of these features, and linearly combines the features to determine the final score used to rank the tuples. 
Given that the original approach was not made publicly available, we implemented L2R. In our implementation, we use a Lucene-based bug localizer \cite{Chaparro2019} 
%to recommend potentially-buggy code files for a particular report, 
as we were unable to obtain the original localizer \cite{Ye2014}. 
%\os{it seems this is not true, so we need to justify this better.}

% \textbf{Lucene} \cite{Hatcher2004} is an IR-based approach that implements a variation of the classical VSM \cite{Salton1975} to compute the textual similarity between documents. We implement an approach where the documents are bug reports. This approach finds the most similar past bug reports to a particular report (the query), 
% %, considering only those submitted within the past six months of report (query) submission. 
% retrieves the developers that addressed the reports, and ranks them in the order in which the reports are ranked (from high to low textual similarity). 

\textbf{Lucene} \cite{Hatcher2004} is an IR-based approach that implements a variation of the classical VSM \cite{Salton1975} to compute the textual similarity between documents. We implemented a Lucene-based approach where the documents are bug reports (similarly to prior duplicate bug report detection work \cite{Runeson2007}). This approach finds the most similar past bug reports to a particular report (the query), retrieves the developers that addressed the reports, and ranks them in the order in which the reports are ranked (from high to low textual similarity).

%\textbf{Lucene} is IR-based approach that uses a variation of the classical VSM \cite{Salton1975}. Inspired by prior duplicate bug report detection research, we defined an approach to find the most similar past bug reports to a particular report (the query) in the first step, then retrieves the developers that addressed the reports, and ranks them in the order in which the reports are ranked (from high to low textual similarity). 

%\cite{Hatcher2004}: for many bug assignment approaches, a prior step is to detect similar bug reports, which is highly related to duplicate bug report detection, thus, following Runeson's paper\cite{Runeson2007}, we implement this IR-based approach that uses a variation of the classical VSM \cite{Salton1975} to find the most similar past bug reports to a particular report (the query) in the first step, then retrieves the developers that addressed the reports, and ranks them in the order in which the reports are ranked (from high to low textual similarity). 

\textbf{Freq} is an approach (inspired by prior work \cite{wang2014fixercache})  that ranks the project developers based on the number of bug reports they have addressed.
% %within the past six months of report submission. 
 Developers that fixed a larger number of reports are ranked higher by this approach.

\vspace{-0.3cm}
\subsection{Metrics and Methodology}
\label{sec:metrics}

%\os{We need to say that the main methodology to answer RQ1 is to determine which is the best performing approach(es) for each bug report and report the percentages (write this nicely)}

%\os{We need to be consistent. Either we use Top-K Accuracy and we use it in the text and tables, or we use H@k (HITS@K)}

We measured the performance of the three approaches using standard metrics used in prior bug assignment studies \cite{Naguib2013,Tian2016a,Sajedi2020}. 
%These are computed on sets of bug reports/queries.

%\textbf{Avg. Rank (AR)} is the average \textbf{\textit{rank}} or position of the first expected developer (\ie from the ground truth) found in the ranked list of developers. The average is computed across all the queries. 
 
 \textbf{HIT@k (H@k)} is the percentage of bug reports (queries) for which at least one of the expected developers (\ie from the ground truth) is found in the top-k recommended developers. We report H@k for $k = 1-5$, as in prior work \cite{Naguib2013, Tian2016a}. \textbf{MRR} is the mean of the queries' reciprocal ranks. The reciprocal rank for a query is 1/\textit{rank}, where \textbf{\textit{rank}} is the position of the first expected developer found in the ranked list of developers. 
%, where \textit{rank} is the position of the 1st expected developer in the list of recommended developers. 
\textbf{MAP} is the mean average precision (AP) over all queries. The AP for a query is the average of the precision values achieved at all the cutting points \textit{k} of the ranked list of developers (precision@k). Precision@k is the proportion of the top-k recommended developers that are correct according to the ground truth.
A higher H@k, MRR, and MAP indicate higher bug assignment performance.
% For example, h-1 is the percentage of ground truth developers ranking first in the list of developers among all queries. 

% We want to evaluate approaches not only by hit rates to the top ranked developers but also the overall performance of recommended list of developers. 
% In addition to hits@k accuracy, we also compute Mean average precision (MAP), Mean reciprocal rank (MRR), precision@k and recall@k (k = 1-5). 
% MAP is the mean of Average Precision, which measures the average accuracy of the retrieval results of all queries.
% MRR is the mean of reciprocal rank, which measures the average of reciprocal of the rank of the first ground truth developer appearing in the retrieval result of all queries. 

%\os{How about the average rank? it is important because we used it to construct the ground-truth for classification.}

%\os{Say here that the higher or lower values the better (e.g., for avg rank, the lower the better - opposite for the other metrics)}

To train/evaluate L2R (as done in \cite{Tian2016a}), 
%we adopt an setting similar to that defined by Tian \etal \re, in which 
we sort the bug reports chronologically by submission date-time and we split them up into 10 ten folds. The first $x$ folds are used for training (with $x=1...9$), and fold $x+1$ is used for model testing. L2R's overall performance is computed with the metrics defined above, applied on the set of queries from all the folds except the first one. The number of queries we used for measuring L2R's performance is 803 (Angular) + 1,208 (WordPress). We measured Lucene's and Freq's performance on the same query set to have a fair comparison among all the approaches.
%, and determine whether there is one approach that performs best for the two systems.

%We also determined the best-performing approach for each individual report/query: 
%, based on \textit{rank}. 
The approach that achieved the lowest \textit{rank} was considered the best approach for each individual report/query. We computed the \% of queries for which each approach performs best, including the cases when the approaches achieved the same (lowest) \textit{rank}.

%We replicate L2R by following this paper and trained a ranking model using our dataset. For the whole 16 activity-based and location-based features, we did not follow the exact method of  generating location-based features due to the unavailability of some information such as xx, instead, we implemented an alternative bug report localization method that can be suitable to our dataset. Following the same experimental setup in the paper, we also sort the bug reports in each benchmark dataset chronologically and split the sorted bug reports into equally folds, and then use the first several folds to train the ranking model and test on the next fold. In this way, for each bug report in test folds, we can test the model by computing the weighted sum of all 16 features of each developer and than rank those developers by their scores, then check if the ground truth appear top-ranked in the list of developers.

\begin{table}[]
\setlength\tabcolsep{4pt}
\small
\centering
\caption{Bug assignment performance for each system}
\label{tab:approach_performance}
\begin{tabular}{clllllll}
\hline
\multicolumn{1}{c|}{\textbf{Approach}} & \textbf{MRR} & \multicolumn{1}{l|}{\textbf{MAP}} & \textbf{H@1} & \textbf{H@2} & \textbf{H@3} & \textbf{H@4} & \textbf{H@5} \\ \hline
\multicolumn{8}{c}{\footnotesize\textbf{Angular}}                                                               \\ \hline
\multicolumn{1}{c|}{Freq}   & 50.7\% & \multicolumn{1}{l|}{43.6\%} & 29.5\% & 49.6\% & 64.3\% & 75.5\% & 82.4\% \\ \hline
\multicolumn{1}{c|}{Lucene} & 40.8\% & \multicolumn{1}{l|}{36.0\%} & 22.0\% & 37.6\% & 46.6\% & 56.4\% & 65.3\% \\ \hline
\multicolumn{1}{c|}{L2R}    & 52.0\% & \multicolumn{1}{l|}{44.7\%} & 31.4\% & 53.0\% & 65.3\% & 74.9\% & 80.5\% \\ \hline
\multicolumn{8}{c}{\footnotesize\textbf{WordPress}}                                                             \\ \hline
\multicolumn{1}{c|}{Freq}   & 49.3\% & \multicolumn{1}{l|}{48.8\%} & 28.5\% & 45.7\% & 60.9\% & 74.4\% & 80.7\% \\ \hline
\multicolumn{1}{c|}{Lucene} & 56.9\% & \multicolumn{1}{l|}{56.2\%} & 38.2\% & 56.8\% & 68.6\% & 77.6\% & 82.7\% \\ \hline
\multicolumn{1}{c|}{L2R}    & 57.0\% & \multicolumn{1}{l|}{56.2\%} & 38.7\% & 56.0\% & 69.4\% & 79.6\% & 84.1\% \\ \hline
\end{tabular}
\end{table}

\begin{table}[]
\centering
\setlength\tabcolsep{2.1pt}
\small
\caption{\% of reports for which the approaches perform best}
\label{tab:query_distribution}
\begin{tabular}{ccccccccc}
\hline
\multicolumn{1}{c|}{\textbf{System}} &
  \textbf{L2R} &
  \textbf{LC} &
  \textbf{FR} &
  \textbf{L2R/LC} &
  \textbf{L2R/FR} &
  \textbf{LC/FR} &
  \multicolumn{1}{c|}{\textbf{All}} &
  \textbf{Total} \\ \hline
\multicolumn{1}{c|}{Angular}   & 21.2\% & 16.7\% & 17.5\% & 4.7\%  & 26.3\% & 4.1\% & \multicolumn{1}{c|}{9.6\%}  & 813   \\ \hline
\multicolumn{1}{c|}{WordPress} & 19.5\% & 20.9\% & 13.4\% & 11.6\% & 9.9\%  & 4.1\% & \multicolumn{1}{c|}{20.6\%} & 1,208 \\ \hline
\multicolumn{9}{c}{\footnotesize LC: Lucene, FR: Freq, All: L2R/LC/FR}                                                   
\end{tabular}
\vspace{-0.5cm}
\end{table}

\subsection{Results}
\label{sec:results}

Table  \ref{tab:approach_performance} shows the bug assignment performance achieved by each approach. The approaches perform differently for each system. For Angular, Freq and L2R achieve a similar performance, while Lucene performs significantly lower. L2R's MRR and MAP are higher than those for Freq mainly because of the higher H@1-3. In terms H@4 and H@5, Freq performs better than L2R. For WordPress, Lucene and L2R outperform Freq on all metrics. Lucene and L2R perform similarly, L2R achieving slightly better MRR/ H@k (except H@2). 
%We conclude that the approaches perform differently for each system.

%Table 1 presents the performance of the three approaches on two projects under the evaluation metrics.
%\os{We need to say here something for meaningful. Explain the table/results (we shouldn't recite the table but explain the results)}
%\os{We need to say that we validated our conjecture.}
%\os{Yes, very important to point out here the implications of the results.}

Table \ref{tab:query_distribution} shows the distribution of bug reports for which the three approaches performed best.
%according to the position (\textit{rank}) of the first expected developer in the recommended developers. 
A different distribution is observed for each system and the distributions are not skewed toward one of the approaches. The proportion of reports for which a single approach performs best ranges from 13.4\% to 21.2\%. For some queries, there are multiple best-performing approaches. For example, for Angular, both L2R and Freq perform best for 26.3\% of the reports, while for WordPress, all the three approaches perform best for 20.6\% of the reports.  For each individual bug report, there is at least one approach that gives a more accurate developer recommendation than the other approaches. This means that no single approach performs equally well for all the reports, thus verifying our conjecture.
%The result motivates us to combine the advantages of the three approaches to automatically suggest the best method for each bug report.

%\os{Table 1 is not the important table to me, Table 2 is the one that shows clearly that for some queries the developer recommendation is better when some approaches are used and not others. So I would place this next paragraph below the next next paragraph.}
%Besides, Table 1 also shows that different approaches perform the best for different projects, that is, no approach perform better than two other approaches on all projects. For example, in terms of the metrics Average\_rank, Freq performs better than other two approaches on Angular.js system, however Lucene performs better for Wordpress-Android. 

%% file: 3_study_2.tex
% !TeX root = 0_main

\section{Study 2}

%\os{We need a better title for this section.} \os{As I mentioned before, maybe we just go with "Study 2" as the title, if we don't have a more creative title}
% \textit{RQ2: What maximum performance can we obtain by combining existing methods ?}
%\os{I like this first paragraph, but maybe we don't need subsection, just an introductory paragraph.}
%\os{Again, I wouldn't say "we propose LUPIN" but rather "we envision an approach that first recommends an approach from the three, that is likely to give a good bug assignment performance, and then apply it. We coined this approach LUPIN..."}

The results from Study 1 motivated us to explore the idea of automatically recommending the best-performing approach for each bug report. In this way, we would combine the advantages of the three approaches to produce better developer recommendations.
%for all the reports in a project. 

We conducted an study to assess the feasibility this idea 
%of recommending bug assignments approaches for individual bug reports 
using machine learning (ML) and 23 features that aim to capture relationships between the bug reports and the bug assignment approaches.
%to recommend the best-performing approach for each report (Sect. \ref{sec:features}). 
ML models recommend the best approach (Step 1), which is applied on the bug report to obtain a ranked list of developers that can address the report (Step 2). 
%To facilitate the presentation of the study, 
We coin this 2-step approach as \textit{Lupin}.

This study aimed to answer the following research question:

%According to the statistics and analysis from the section 2.4, We can see the different approaches performs differently among different queries, which implies a huge potential to obtain better performance by combining the three approaches. An intuitive solution of combining three approaches is to design a classifier to predict the best-performing approach according to the features of bug reports. 

\begin{center}
    \textit{\textbf{RQ$_2$}: What is the bug assignment performance of Lupin compared to that of baseline bug assignment approaches?} 
    %How can we improve the bug triaging performance using existing methods?
\end{center}

\subsection{Model Features}
\label{sec:features}
%\os{We need to cite the papers that inspired us to define these metrics (Sonia Haiduc's papers).}

%\os{We need to emphasize this better. I would put this relationship as the first sentence of the paragraph.}.

We used 23 features that aim to capture possible relationships between the bug reports and the bug assignment approaches. We selected 18 features from prior work on query quality assessment \cite{Mills2017} since these can be used to measure textual and statistical properties of the bug report text and the textual corpora of past bug reports and past code changes made by the developers, information that is leveraged by L2R and Lucene. We defined additional five (5) features to capture the bug fixing distribution of developers in a project, information that is leveraged by Freq. We summarize the features.

%We expect the ML models to learn relationships among the features to recommend the best approach for a bug report.

%The 18 features selected from prior work \cite{Mills2017} fall into three categories. 
\textit{Specificity} features (11/23) measure how specific the bug report text is compared to the corpus of past bug reports (\aka documents) \cite{Mills2017} to differentiate
%. If the report terms commonly exist in the corpus reports, it may be hard to differentiate 
relevant and non-relevant documents. One specificity feature is the standard deviation of the inverse document frequency of the bug report terms (stdev. IDF). If the stdev. IDF for a bug report is low, L2R's and Lucene's recommendations may be impacted because the bug report may not have discriminatory information to identify the relevant past reports.

\textit{Similarity} features (6/23) measure the degree of textual similarity between the query and the corpus of past bug reports or code files modified by the developers \cite{Mills2017}, information used by Lucene and L2R, respectively. Higher similarity may indicate the existence of many relevant documents to the report/query \cite{Mills2017}, potentially leading to better developer recommendations. One feature is the avg. Collection Query Similarity: the linear combination between the frequency of a report term in the collection of past code files and its IDF. % The avg. (over the query terms) 
This is computed by averaging over all query terms.

%It should be easier to retrieve similar documents if the query is similar with the corpus. \os{What's the difference between similarity and coherency?}
We used one (1/23) \textit{coherency} feature that measures the average similarity between pairs of past bug reports that contain a bug report term \cite{Mills2017}. This feature measures how focused a bug report is on a particular topic, expressed by its vocabulary. This information is used by both Lucene and L2R.

%the level of inter-similarity between the corpus documents containing the bug report terms, which may lead to difficulty in identifying the similar documents if the query is more coherent. \os{I don't understand what coherency means.}

The remaining five (5/23) features measure the bug fixing activity of the developers. For example, some of these features measure the average/median/maximum \# of reports solved by the developer, and the \# of active developers (information leveraged by Freq).

\subsection{Models and Methodology}
\label{sec:classification_methodology}

%\os{Let's compute the following: \% of bugs in the whole data for each approach, \% of bugs in the training data for each approach, and \% of bugs in the testing data for each approach}
%\os{We need to describe here how we built the ground-truth data for classification. And we need to show statistics about the data. For example, the proportion of bugs for each class. Maybe we need to discuss how we dealt with data imbalance?}
%\yang{we did not deal with data imbalance stuff, it is obvious we have data imbalance here, because we choose freq as the first choice if we have same-performance approaches, so we think we ignore this problem here and do not complicate things }

%Automated approach recommendation is formulated as a multi-class classification problem where each bug report is assigned to one of three classes (approaches).
%In other words, one approach is recommended for a single report, one that is more likely to give more accurate recommendations.
We used well-known classifiers/models, used in prior bug assignment research \cite{Bhattacharya2012}, namely Random Forest, Naive Bayes, Decision Trees, and Logistic Regression, to automatically assign a bug report to one of three classes (\ie the approaches).

To train/evaluate the  classifiers, we built ground-truth data based on the results presented in Table \ref{tab:query_distribution}. For each bug report, we selected the approach with the lowest \textit{rank} as the ground-truth approach for the report. In case of \textit{rank} ties, we opted for randomly selecting Freq, Lucene, or L2R as the ground-truth approach, to avoid data imbalance in our classes by selecting a single approach (as we tried to do in pilot experiments), which would potentially bias the classifiers. To address potential effects of the randomness, we repeat the experimental setting described in this subsection five times (\ie having five ground truth datasets).
%In case of ties among approaches, we selected either Freq, Lucene, or L2R in that order as the ground-truth approach. With this strategy, we give more priority to the simpler approaches, which may be more practical to be used in practice. 

%Why random?
% 1. If select one specific approach we would get imbalanced data, which could bias the models
% 2. In principle, any approach could be selected as best.

%Our adoption of these techniques are because they are widely used and proven to be effective in bug triage in software engineering \os{Do they have to be used in bug triage? I don't think so. Maybe we need to justify the models based on the size of our data (\# of features and instances)? Also, add citations here.}.\yang{I remembered these classifiers are widely used in bug triage papers, I can put some papers here}
To train the classifiers, we first sort the bug reports by submission date-time for each system. Next, we take the first 70\% for training/validation, and the remaining 30\% for testing. We use 5-fold cross validation to select the best parameters of each classifier using a chronological splitting where the first $x$ folds (with $x=1...4$) are used for training and the $x+1$ fold is used for validation. 
%On aggregate, the ground-truth reports have L2R, Lucene, and Freq as the best approach in 20.2\%, 28.1\%, and 51.8\% of the cases. The training/validation and test sets have a similar report distribution.

We found the best parameters for each classifier as those that led to the largest weighted avg. \textit{F1-score} over the three classes on the validation sets.
%, computed based on the proportion of reports for each class \re. 
%The F1-score is the harmonic mean between \textit{precision} and \textit{recall} for class~$c$. 
%Precision is the proportion of reports that are assigned to class $c$ (by a model) and are correct according to the ground-truth. Recall is the proportion of reports in class $c$ (according to the ground-truth) that are assigned correctly to class~$c$. 
We measured the classification performance of each classifier (with the best parameters) on the test set using weighted avg. \textit{F1-score}, \textit{precision}, and \textit{recall}. We average these values over the five runs for each classifier to obtain an overall performance

%We use the same metrics for measuring the classification performance of each classifiers on the test set 
%(244 reports for Angular and 362 for WordPress) 
%and report the weighted avg. of the metrics over the three classes, averaged over the five runs. 
%The best performing ML model is the one with the highest weighted average  F1-score, which means that we consider precision and recall equally important.

We experimented with the four classifiers by running them in \textit{Lupin}'s 1st step. Then, in the 2nd step, the predicted approach (given by the classifier) for each report was executed to obtain a list of developers. \textit{Lupin}'s performance is measured using the metrics from Sect. \ref{sec:metrics} on the test set when using each classifier. To compute the overall \textit{Lupin}'s performance, we averaged these metrics over the five runs. We report \textit{Lupin}'s best overall performance by selecting the classifier that leads to the lowest MRR, as this metric might better capture the scenario where the bug triager scans through the developer list until deciding the developer for the bug report.

\begin{table}[]
\caption{Bug assignment performance on the test set}
\label{tab:lupin_results}
\setlength\tabcolsep{4pt}
\small
\begin{tabular}{clllllll}
\hline
\multicolumn{1}{c|}{\textbf{Approach}} & \textbf{MRR} & \multicolumn{1}{l|}{\textbf{MAP}} & \textbf{H@1} & \textbf{H@2} & \textbf{H@3} & \textbf{H@4} & \textbf{H@5} \\ \hline
\multicolumn{8}{c}{\footnotesize\textbf{Angular}}                                                                       \\ \hline
\multicolumn{1}{c|}{L2R}            & 56.9\% & \multicolumn{1}{l|}{49.9\%} & 32.8\% & 61.9\% & 79.9\% & 87.7\% & 90.6\% \\ \hline
\multicolumn{1}{c|}{Lucene}         & 43.8\% & \multicolumn{1}{l|}{39.0\%} & 21.5\% & 40.5\% & 54.9\% & 67.5\% & 78.1\% \\ \hline
\multicolumn{1}{c|}{Freq}           & 57.4\% & \multicolumn{1}{l|}{50.2\%} & 34.9\% & 57.6\% & 76.5\% & 89.5\% & 93.7\% \\ \hline
\multicolumn{1}{c|}{\textit{Lupin}} & 56.5\% & \multicolumn{1}{l|}{48.9\%} & 32.8\% & 59.7\% & 77.0\% & 88.4\% & 92.2\% \\ \hline
\multicolumn{1}{c|}{\textit{Max}}   & 72.2\% & \multicolumn{1}{l|}{62.6\%} & 56.6\% & 75.8\% & 86.5\% & 91.4\% & 94.7\% \\ \hline
\multicolumn{8}{c}{\footnotesize\textbf{WordPress}}                                                                     \\ \hline
\multicolumn{1}{c|}{L2R}            & 46.0\% & \multicolumn{1}{l|}{45.3\%} & 29.0\% & 41.7\% & 54.0\% & 62.4\% & 68.7\% \\ \hline
\multicolumn{1}{c|}{Lucene}         & 47.8\% & \multicolumn{1}{l|}{47.2\%} & 30.7\% & 44.3\% & 53.3\% & 61.4\% & 68.1\% \\ \hline
\multicolumn{1}{c|}{Freq}           & 40.4\% & \multicolumn{1}{l|}{40.1\%} & 23.4\% & 30.8\% & 41.9\% & 53.3\% & 63.2\% \\ \hline
\multicolumn{1}{c|}{\textit{Lupin}} & 46.2\% & \multicolumn{1}{l|}{45.7\%} & 29.1\% & 41.1\% & 52.7\% & 61.8\% & 68.4\% \\ \hline
\multicolumn{1}{c|}{\textit{Max}}   & 63.7\% & \multicolumn{1}{l|}{62.8\%} & 48.3\% & 61.8\% & 74.1\% & 81.3\% & 87.6\% \\ \hline
\end{tabular}
\vspace{-0.5cm}
\end{table}

\begin{comment}

\begin{table}[]
\caption{Bug assignment performance on the test set}
\label{tab:lupin_results}
\setlength\tabcolsep{3pt}
\small
\begin{tabular}{cllllllll}
\hline
\textbf{Approach} & \textbf{AR}  & \textbf{MRR}    & \textbf{MAP}    & \textbf{H@1}    & \textbf{H@2}    & \textbf{H@3}    & \textbf{H@4}    & \textbf{H@5}    \\ \hline
\multicolumn{9}{c}{\footnotesize\textbf{Angular}}                                                   \\ \hline
L2R      & 3.0 & 56.9\% & 49.9\% & 32.8\% & 61.9\% & 79.9\% & 87.7\% & 90.6\% \\ \hline
Lucene   & 4.0 & 43.8\% & 39.0\% & 21.5\% & 40.5\% & 54.9\% & 67.5\% & 78.1\% \\ \hline
Freq     & 2.7 & 57.4\% & 50.2\% & 34.9\% & 57.6\% & 76.5\% & 89.5\% & 93.7\% \\ \hline
\textit{Lupin}    & 2.8 & 56.5\% & 48.9\% & 32.8\% & 59.7\% & 77.0\% & 88.4\% & 92.2\% \\ \hline
\textit{Max}     & 2.2 & 72.2\% & 62.6\% & 56.6\% & 75.8\% & 86.5\% & 91.4\% & 94.7\% \\ \hline
\multicolumn{9}{c}{\footnotesize\textbf{WordPress}}                                                 \\ \hline
L2R      & 5.3 & 46.0\% & 45.3\% & 29.0\% & 41.7\% & 54.0\% & 62.4\% & 68.7\% \\ \hline
Lucene   & 4.3 & 47.8\% & 47.2\% & 30.7\% & 44.3\% & 53.3\% & 61.4\% & 68.1\% \\ \hline
Freq     & 4.7 & 40.4\% & 40.1\% & 23.4\% & 30.8\% & 41.9\% & 53.3\% & 63.2\% \\ \hline
\textit{Lupin}    & 4.7 & 46.2\% & 45.7\% & 29.1\% & 41.1\% & 52.7\% & 61.8\% & 68.4\% \\ \hline
\textit{Max}     & 3.0 & 63.7\% & 62.8\% & 48.3\% & 61.8\% & 74.1\% & 81.3\% & 87.6\% \\ \hline
\end{tabular}
\vspace{-0.5cm}
\end{table}
\end{comment}

\subsection{Results and Discussion}
\label{sec:lupin_results}
%\os{Move this RQ above and rephrase it to read clearer and reflect better the goal of the study}

%\os{This is the most important subsection of the paper. I read it completely and it reads well. It is just that we need to improve the writing and sound a little more convincing.}

%Table \ref{tab:classification_results} shows the classification performance of the models on the test set. Random Forest (RF) and Naive Bayes (NB) achieve the highest F1 score (49.1\% and 38.9\%) for Angular and  WordPress, respectively. These are the models that are used with \textit{Lupin}. Decision Trees (DT) achieves a similar performance to NB for WordPress, but achieves the lowest performance for Angular. The models' performance is not very high, which indicate that the models may require more information to distinguish among the approaches. One potential solution we will explore in future work is to use post-retrieval features \re rather than only pre-retrieval features.

%Table 3 shows the results of the classifiers. For Angular.js system, the best classifier is Decision Tree, which achieves 42.42\% F1 score, so we select it as a selector in LUPIN. For Wordpress-Android system, the best classifier is Logistic Regression which is selected as the selector in LUPIN in terms of F1 score. 

Table \ref{tab:lupin_results} shows the bug assignment performance of \textit{Lupin} (on the test set) compared to that of the baseline approaches (L2R, Lucene, \& Freq). 
The \textit{max} row in Table  \ref{tab:lupin_results} shows the maximum performance that Lupin can achieve if it used a classifier that perfectly predicts the best approach for every bug report. The \textit{max} results indicate that there is great potential for Lupin to improve bug assignment performance, yet the \textit{Lupin} results show that it is still far from achieving the maximum performance, as are the baseline approaches.

For Angular, \textit{Lupin} (using Naive Bayes) achieves comparable performance to L2R/Freq's performance and superior performance to Lucene's. Similarly, for WordPress, \textit{Lupin} (using Decision Trees)  achieves comparable performance to L2R/Lucene's performance and superior performance to Freq's. The results show that \textit{Lupin} does not perform more accurately than all the baseline approaches. 

We investigated this phenomenon by analyzing the overall classification performance of the classifiers on the test set. We found that Naive Bayes achieves 30.6\%, 36.7\%, and 31.3\% weighted avg. precision, recall, and F1, respectively, while Decision Trees achieve 32.9\%, 34.4\%, and 29.3\% weighted avg. precision, recall, and F1, respectively. These are average values over the five experimental runs. It is important to note that the \% of reports for each class (approach) range from 23\% (min) to 40\% (max) across the five runs, which rules out potential problems related to data imbalance. One likely reason for the classifier results is that the classifier features are not capturing enough information to distinguish among the approaches. One potential solution to explore in our future work is to use post-retrieval features \cite{Mills2017} rather than only pre-retrieval features (as the ones we currently used).

However, since the features we used for classification encode information that L2R, Lucene, and Freq leverage to make developer recommendations, one potential implication of the results is that these (and other) approaches may also lack additional information and factors that maintainers use in practice to assign bug reports to developers, that are not necessarily found in software repository data (in bug reports, commits, source code, \etc) \cite{Aranda2009}. For example, maintainers may assign reports to developers based on developer availability at a given moment, developer's technical background or experience (\eg in using specific technology), and different social structures and dynamics that can be found in open source projects \cite{bird2008latent,zou2018practitioners}. In other words, maintainers may perform bug assignment not entirely based on developer profiles defined by the past bug reports, code changes that developers have addressed, and other information, as existing bug assignment techniques attempt to model. We advocate for additional research to understand the way bug reports are assigned to developers in practice.

%% file: 5_threats.tex
\section{Threats to Validity}

%\os{We don't need subsections here, it is enough to have one or two paragraphs discussing the main threats and what we did to mitigate them. Just writing here some ideas, but of course need some expansion.}
%\os{Internal Validity: Selection of systems and bug assignment approaches, ground-truth creation for classification, ground-truth creation for bug assignment}

%\os{Construct validity: the features may not capture entirely well the relationships between bugs and approaches?}

%\os{Conclusion validity: maybe our interpretation of the RQ2 results (i.e., the implications) is not entirely correct?}

%\os{External Validity: we tried to be diverse on selecting the systems/approaches, but future work will use more systems and approaches.}

%We discuss the most important threats to validity. 
%While we used methodological decisions and metrics used in prior bug assignment research to design our empirical studies, we are not exempt of threats to validity. 
%They way we created ground-truth data is one validity threat. 
As in prior work \cite{Sajedi2020}, we used project repository data to create bug assignment ground-truth data. We minimized potential errors in bug reports and developer sets by (1) selecting issues labeled as ``bugs'' by the original maintainers, and (2) by manually curating the extracted data. For example, we manually merged GitHub accounts that referred to the same developer. 
%The implementation of approaches, especially L2R \re, is another threat. 
We selected a diverse set of approaches according to the techniques they implemented and the information they used. 
%We attempted to replicate/implement L2R , however, we did not use the original bug localizer because we were not able to obtain it. We 
%The selected features may not entirely capture possible relationships between bug reports and the three approaches.
The results may not generalize beyond the selected two open source projects. Expanding the studies with additional approaches/systems is in our plans for future work.

%% file: 6_related_work.tex
\section{Related work}

A variety of approaches have been proposed to automatically assign bug reports to developers
\cite{aktas2020automated,Anvik2006,Anvik2011,baysal2009bug,Bhattacharya2012,bortis2013porchlight,hossen2014amalgamating,Jeong2009,jonsson2016automated,Kagdi2012,Kevic2013,lee2017applying,linares2012triaging,Matter2009,Naguib2013,sajedi2016crowdsourced,Sajedi2020,sarkar2019improving,Shokripour2013,shokripour2015time,sun2017enhancing,Tamrawi2011a,Tamrawi2011b,Tian2016a,xia2015dual,Xia2017,zanjani2015using,Zhang2017,zhang2020efficient}.
%\cite{aktas2020automated,Anvik2006,Anvik2011,baysal2009bug,Bhattacharya2012,bortis2013porchlight,hossen2014amalgamating,Jeong2009,jonsson2016automated,Kagdi2012,Kevic2013,lee2017applying,linares2012triaging,Matter2009,Naguib2013,sajedi2016crowdsourced,Sajedi2020,sajedi2020vocabulary,sarkar2019improving,Shokripour2013,shokripour2015time,sun2017enhancing,Tamrawi2011a,Tamrawi2011b,Tian2016a,Wu2011,Xia2013,xia2015dual,Xia2017,xie2012dretom,xuan2017automatic,zanjani2015using,Zhang2017,zhang2020efficient}.
%, which range from machine learning models \re and information retrieval techniques \re to topic models \re and tossing-graph-based techniques \re, among others \re. 
These techniques leverage multiple sources of information such as past bug reports, source code, and bug tossing information, and are typically designed to be applied to all bug reports in a project. In contrast, our work aims to recommend specific approaches to individual bug reports to 
%taking advantage of the benefits of existing approaches and the nuances of each bug report to perform 
improve developer recommendations. To the best of our knowledge, this work is the first to investigate this idea, yet similar work has been done to support other software engineering tasks \cite{Mills2017,moreno2015query,chen2021my}.
%The work most closely related to ours is by Mills \etal \re, who explored how to select query reformulation methods (via machine learning) for a specific bug report in the context of feature/bug localization.

%Many approaches have been proposed in the literature to automatically assign a given bug report to the appropriate developers\cite{murphy2004automatic, Anvik2006, Jeong2009, Matter2009, Anvik2011, Naguib2013, wang2014fixercache, Tian2016a, Zhang2017, Sajedi2020}. Similar with our work, most bug assignment approaches use different artifact information (e.g, textual information and metadata such as bug reports\cite{Runeson2007, Sun2011}, commits, source code\cite{linares2012triaging}, developers, version, priority, etc.) to encode such information for recommending suitable developers. However, existing approaches did not consider the nuances of different approaches, thus, based on our finding that these approaches may not work equally well for all the reports in one software project, our work is intended to recommend one suitable approach rather than developers.

%% file: 7_conclusions.tex
% !TeX root = 0_main

\section{Conclusions and Future Work}

%\os{Report the findings and implications}

We conducted an empirical study that applied three bug assignment approaches on 2k+ bug reports from two open source projects. We found that such approaches do not perform equally well when applied on all the bug reports from a software project. %This means that they work better for some reports and worse for others. 
This finding motivated us to explore the idea of automatically recommending and applying the best-performing approach on individual bug reports via machine learning (ML). We experimented with four ML models that learn from 23 features and found that this composite approach is far from achieving the maximum possible performance, while achieving comparable performance to that of the baselines approaches. We found that the features utilized by the ML models may not capture enough information to distinguish between approaches. A possible implication of this result is that bug assignment approaches do not capture factors (\eg developer availability) found in the way bug reports are assigned to developers in practice. The results warrant future research in (1) defining effective features to better distinguish assignment approaches, (2) understanding how developers perform bug assignment in practice, and (3) incorporating additional information on automated bug assignment approaches to better recommend developers.

%This paper firstly presents an empirical study that explore how bug report assignment approaches perform and compare when applied on individual bug reports. Our empirical evaluation, conducted on two open-source projects using three bug assignment approaches, verified our conjecture that existing approaches may not work well equally for all the bug reports in one software project. This finding motivates us to implement a preliminary study to explore the idea of designing a composite approach that automatically recommend the best approach from three existing approaches using machine learning. The result shows that our initial attempt is still far from achieving the maximum possible performance, we discuss possible reasons for these results which warrants more future work in explore how developers assign bug reports and then improve the existing approaches. Our future work will focus on (1) exploring more effective features that can capture enough information that distinguish different approaches; (2) exploring the limitation of existing bug assignment approaches by understanding how developers assign bugs in practice via the format of survey, which in return may help us design more effective features for bug assignment. 

%% file: reference.bbl
%%% -*-BibTeX-*-
%%% Do NOT edit. File created by BibTeX with style
%%% ACM-Reference-Format-Journals [18-Jan-2012].